\documentclass[12pt]{article}
\pdfoutput=1
\usepackage[latin1]{inputenc}
\usepackage{amsmath}
\usepackage{amsfonts}
\usepackage{amssymb}
\usepackage{graphicx}
\usepackage{geometry}
\usepackage{amssymb,epsfig}
\usepackage{hyperref}
\usepackage{subfig}

\makeatletter
\renewcommand\section{\@startsection {section}{1}{\z@}%
                                 {-3.5ex \@plus -1ex \@minus -.2ex}
                                   {2.3ex \@plus.2ex}%
                                   {\normalfont\large\bfseries}}
\renewcommand\subsection{\@startsection{subsection}{2}{\z@}%
                                   {-3.25ex\@plus -1ex \@minus -.2ex}%
                                     {1.5ex \@plus .2ex}%
                                     {\normalfont\bfseries}}
\renewcommand\subsubsection{\@startsection{subsubsection}{3}{\z@}%
                                   {-3.25ex\@plus -1ex \@minus -.2ex}%
                                     {1.5ex \@plus .2ex}%
                                     {\normalfont\itshape}}
\makeatother

\def\pplogo{\vbox{\kern-\headheight\kern -29pt
\halign{##&##\hfil\cr&{\ppnumber}\cr\rule{0pt}{2.5ex}&\ppdate\cr}}}
\makeatletter
\def\ps@firstpage{\ps@empty \def\@oddhead{\hss\pplogo}%
  \let\@evenhead\@oddhead 
}
\def\maketitle{\par
 \begingroup
 \def\thefootnote{\fnsymbol{footnote}}
 \def\@makefnmark{\hbox{$^{\@thefnmark}$\hss}}
 \if@twocolumn
 \twocolumn[\@maketitle]
 \else \newpage
 \global\@topnum\z@ \@maketitle \fi\thispagestyle{firstpage}\@thanks
 \endgroup
 \setcounter{footnote}{0}
 \let\maketitle\relax
 \let\@maketitle\relax
 \gdef\@thanks{}\gdef\@author{}\gdef\@title{}\let\thanks\relax}
\makeatother

\numberwithin{equation}{section}

\newcommand{\be}{\begin{equation}}
\newcommand{\bea}{\begin{eqnarray}}
\newcommand{\ee}{\end{equation}}
\newcommand{\eea}{\end{eqnarray}}
\newcommand\beq{\begin{equation}}
\newcommand\eeq{\end{equation}}

\newcommand{\mc}{\mathcal}

\renewcommand{\t}{\tilde}

\begin{document}

\setcounter{page}0
\def\ppnumber{\vbox{\baselineskip14pt
}}
\def\ppdate{\footnotesize{SLAC-PUB-14850 ~~~ SU-ITP-12/01}} \date{}

\author{Xi Dong, Sarah Harrison, Shamit Kachru, Gonzalo Torroba and Huajia Wang\\
[7mm]
{\normalsize \it Stanford Institute for Theoretical Physics }\\
{\normalsize  \it Department of Physics, Stanford University}\\
{\normalsize \it Stanford, CA 94305, USA}\\
[7mm]
{\normalsize \it Theory Group, SLAC National Accelerator Laboratory}\\
{\normalsize \it Menlo Park, CA 94309, USA}\\
}

\bigskip
\title{\bf  Aspects of holography for theories  
with hyperscaling violation
\vskip 0.5cm}
\maketitle

\begin{abstract}

We analyze various aspects of the recently proposed holographic theories with general dynamical critical exponent $z$ and hyperscaling violation exponent $\theta$.
We first find the basic
constraints on $z,\theta$ from the gravity side, and compute the stress-energy
tensor expectation values and scalar two-point functions. Massive correlators exhibit a nontrivial exponential behavior at long distances, controlled by $\theta$. At short distance, the two-point functions become power-law, with a universal form for $\theta > 0$. Next, the calculation of the holographic entanglement entropy reveals the existence of novel phases which violate the area law. The entropy in these phases has a behavior that interpolates between that of a Fermi surface and that exhibited by systems with extensive entanglement entropy.
Finally, we describe microscopic embeddings of some $\theta \neq 0$ metrics into full string theory models -- these metrics characterize
large regions of the parameter space of Dp-brane metrics for $p\neq 3$.  For instance, the theory of $N$ D2-branes in IIA
supergravity has $z=1$ and $\theta = -1/3$ over a wide range of scales, at large $g_s N$.
\end{abstract}
\bigskip
\newpage

\tableofcontents

\vskip 1cm

\section{Introduction}\label{sec:intro}

Holography \cite{tS} is a powerful tool to study strongly interacting large $N$ quantum field theories \cite{Juan,GKP,Witten,Holography}.  In the holographic context, a 
$d$ (spatial) dimensional quantum field theory is mapped to a higher-dimensional (usually $(d+2)$-dimensional) gravitational theory, with the $(d+1)$
``field theory dimensions'' arising as the boundary of the space-time.  While the initial interest in concrete examples centered on applications to
AdS gravity theories and their conformal field theory duals, the class of metrics of interest in gauge/gravity duality has been considerably enlarged
in recent years.  One simple generalisation is to consider metrics which can be dual to scale-invariant field theories which are, however, ${\it not}$
conformally invariant, but instead enjoy a dynamical critical exponent $z \neq 1$ (with $z=1$ reducing to the case of the AdS metric):
\begin{equation}
\label{Lifz}
ds^2 = -\frac{1}{r^{2z}} dt^2 + \frac{1}{r^2} (dr^2 + dx_i^2) ~ .
\end{equation}
These metrics are invariant under the scaling
\begin{equation}
\label{scaling}
t \to \lambda^z t,~ x_i \to \lambda x_i, ~r \to {\lambda r }~.
\end{equation}
They arise as exact solutions of simple gravity theories coupled to appropriate matter \cite{KLM, Taylor}, with the simplest such theory
also including an abelian gauge field in the bulk.
This simple generalisation  of AdS is motivated by consideration of gravity toy models of condensed matter systems (where Lorentz invariance
needn't emerge in the deep infrared, and e.g. doping with charge density can naturally lead to $z \neq 1$).\footnote{An alternative class of non-relativistic scaling metrics characterizing the so-called Schr\"odinger space-times, which may also have condensed matter applications, was discussed in \cite{Schrodinger}.}
Such metrics have also been found as 
solutions in string theory, and supergravities which arise simply from string constructions, in \cite{Lifembeddings}.

More recently, it has been realized that by studying systems including a scalar ``dilaton'' in the bulk, one can find even larger classes of
scaling metrics.   Such theories have been studied in e.g. \cite{Gubser,GKPT,Cadoni,Kiritsis,Eric,Peet,GKPTIW,Sandip,Berglund,Ogawa:2011bz,Huijse:2011ef,Shaghoulian:2011aa} (with very similar metrics also characterizing the ``electron
stars'' studied in e.g. \cite{Sean}).
By including both an abelian gauge field and the scalar dilaton,  one can in particular engineer the full class of metrics \cite{Kiritsis}
\begin{equation}\label{thetaz}
ds_{d+2}^2 = r^{-2(d-\theta)/d} \left( - r^{-2(z-1)} dt^2 + dr^2 + dx_i^2 \right)~.
\end{equation}
These exhibit both a dynamical critical exponent $z$ and a ``hyperscaling violation'' exponent $\theta$ \cite{Fisher}, as emphasized
in \cite{Huijse:2011ef}.  This metric is not scale invariant, but transforms as
\begin{equation}
ds \to \lambda^{\theta/d} ds
\end{equation}
under the scale-transformation (\ref{scaling}).
Very roughly speaking, in a theory with hyperscaling violation, the thermodynamic behaviour is as if the theory enjoyed dynamical exponent $z$ but
lived in $d-\theta$ dimensions; dimensional analysis is restored because such theories typically have a dimensionful scale that \textit{does not decouple
in the infrared}, and below which
such behaviour emerges.  One can then use appropriate powers of this scale -- which will be denoted by $r_F$ -- to restore naive dimensional analysis.  As emphasized in \cite{Huijse:2011ef}, building on 
the stimulating explorations in \cite{Ogawa:2011bz}, the case $\theta = d-1$ is a promising gravitational representation of a theory with a Fermi surface
in terms of its leading large $N$ thermodynamic behaviour.  In this
example, the relevant dimensionful scale is of course the Fermi momentum.

In this paper we characterize strongly coupled quantum field theories with hyperscaling violation using holography.
In general, the metric (\ref{thetaz}) may not be a good description above the dynamical scale $r_F$.\footnote{For instance, in systems controlled by a UV fixed point, the metric should become scale invariant for $r \ll r_F$. Similarly, the hyperscaling violation regime may also be changed in the deep IR if there is an attractive fixed point. We will find string theory examples of these situations.} For this reason, in this work we will not assume (\ref{thetaz}) all the way to the boundary, but instead we will follow an `effective' holographic approach in which the dual theory lives on a finite $r$ slice. This is similar to an effective field theory analysis in the dual QFT. This has been put on a firmer footing for asymptotically AdS spacetimes in~\cite{RG}.

First, we discuss the most basic holographic features of this class of metrics: the constraints on $(z,\, \theta)$ that follow from energy conditions in the
bulk, the behavior of propagators for bulk fields (and the consequent behavior of correlation functions of scalar operators in the dual field theories), and 
the behavior of the stress-energy tensor. Our analysis reveals intriguing properties of correlation functions in these theories. In a semiclassical approximation, a massive scalar has a correlation function of the form 
\be
G(\Delta x) \sim \exp \left(- c_\theta \, \frac{m}{\theta} |\Delta x|^{\theta/d} \right)
\ee
at spacelike separations (where $c_\theta>0$ is a constant). We note the nontrivial $|\Delta x|^{\theta/d}$ dependence, as compared to a weakly coupled massive propagator, $G(\Delta) \sim \exp(-m |\Delta x|)$. On the other hand, away from the semiclassical limit (i.e. at small masses/short distances), there is a cross-over to a power-law behavior, and the propagator becomes\footnote{This assumes $\theta>0$; see \S \ref{sec:scalars} for more details.}
\be
G(\Delta x) \sim \frac{1}{|\Delta x|^{2(d+1)-\theta}}\,.
\ee
That is, there is a universal $\theta$-dependent power law, independent of $m$.
 
In another direction, we systematically
study the entanglement entropy properties of the dual field theories.
In recent years, studies of entanglement entropy have come to the fore as a new technique for understanding and perhaps classifying novel phases
of quantum field theory \cite{EEreview}.  
In general, the entanglement between a region $A$ and its complement in a quantum field theory in its ground state in $d$ spatial dimensions, is expected to
scale as the
${\it area}$ of $\partial A$, the boundary of the region (with a precise proportionality constant dependent on the UV-cutoff of the field theory) \cite{Srednicki}.
For the UV-dependent contribution, this scaling simply follows from locality, and has come to be known as the ``area law.''  However, 
several states which violate the area law have also been discussed.  These include $d=1$ conformal field theories \cite{Wilczek}, conventional Fermi liquids, which can exhibit logarithmic
violation of the area law \cite{Fermions,SwingleFS}, and certain proposed non-Fermi liquid ground states \cite{NFL}.   In these systems,
the area law - violating terms are ${\it not}$ cut-off dependent.
More generally, sub-leading but cut-off independent terms in the entanglement entropy have proved to be of 
significant interest -- for instance, they can distinguish between states with
different topological orders \cite{KitaevPreskill,LevinWen}.   

Ryu and Takayanagi \cite{Ryu:2006bv} proposed that the entanglement entropy between a region A and its complement in the
boundary field theory, can be computed in gravity by finding the area of the minimal surface ending on $\partial A$ and
extending into the bulk geometry (in Planck units).  While this proposal is as yet unproven, it has passed many non-trivial tests, and is
supported by an impressive amount
of circumstantial evidence.
Here, we systematically study the entanglement properties of the class of metrics (\ref{thetaz}), over the full range of parameters
$z, \theta$ where they seem to emerge as solutions of a reasonable matter + gravity theory (i.e., one where the required stress-energy sources satisfy reasonable energy conditions).
Entanglement properties of subsets of these theories were studied in \cite{Ogawa:2011bz,Huijse:2011ef,Shaghoulian:2011aa}, and also in \cite{Swingle}, which emphasized the importance
of the cross-over between the area law at $T=0$ and the thermal entropy.  One of the surprises
we'll encounter is the existence of a class of theories which violate the area law and have universal terms in the ground-state holographic entanglement entropy that
scale parametrically faster than the area of $\partial A$ (while scaling more slowly than the extensive entanglement entropy expected in a theory with extensive ground-state
entropy \cite{Swingle}).

As a third focus, we also discuss the way some $\theta \neq 0$ metrics arise in a UV-complete theory -- string theory.  Existing embeddings have been in
phenomenological theories of Einstein/Maxwell/dilaton gravity, which are clearly applicable only over some range of scales (as the dilaton is varying, leading one
to suspect that the description breaks down both in the deep IR and the far UV).  Here, we simply point out that over a wide range of scales, the dilatonic Dp-branes
(those with $p \neq 3$) give rise to metrics of the form (\ref{thetaz}) with $z=1$ but $\theta \neq 0$.\footnote{After submitting this work, we learned that a discussion which overlaps with our own discussion of non-conformal D-branes appears in \cite{Eric}.}  
For instance, the 
large N D2-brane theory, in the IIA supergravity regime, has $\theta = -1/3$.  The string embedding, together with our knowledge of the properties of Dp-branes, 
provides a complete understanding of what happens in the far UV and deep IR
regions of the phase diagram where ``bottom up'' descriptions break down \cite{IMSY}.

\section{Holographic theories with hyperscaling violation}\label{sec:holography}

We begin by analyzing basic properties of theories with hyperscaling violation using holographic techniques. In this first step, our goal will be to determine two-point functions and the expectation value of the energy-momentum tensor for these field theories. In the following sections we will construct other observables, such as the entanglement entropy, and study finite temperature effects.

\subsection{Metrics with scale covariance}\label{subsec:covariant}

As we reviewed before, the gravity side is characterized by a metric of the form
\be\label{eq:dsIR1}
ds^2_{d+2} =  r^{-2(d-\theta)/d} \left(-r^{-2(z-1)} dt^2 + dr^2 + dx_i^2 \right)\,.
\ee
This is the most general metric that is spatially homogeneous and covariant under the scale transformations
\be
x_i \to \lambda x_i\;,\;t \to \lambda^z t\;,\;r \to \lambda r\;,\;ds \to \lambda^{\theta/d} ds\,.
\ee
Thus, $z$ plays the role of the dynamical exponent, and $\theta$ is the hyperscaling violation exponent.

The dual $(d+1)$-dimensional field theory lives on a background spacetime identified with a surface of constant $r$ in (\ref{eq:dsIR1}). The radial coordinate is related to the energy scale of the dual theory. For example, an object of fixed proper energy $E_{pr}$ and momentum $\vec{p}_{pr}$ redshifts according to
\be
E(r) = \frac{1}{r^{z-\theta/d}} E_{pr}\;,\; \vec{p}(r) = \frac{1}{r^{1-\theta/d}}\vec{p}_{pr}\,.
\ee
When $\theta \le d z$ and $\theta<d$, $r \to 0$ (the boundary of (\ref{eq:dsIR1})) describes the UV of the dual QFT.
Clearly, different probes give different energy-radius relations, as in AdS/CFT~\cite{Peet:1998wn}. For instance, a string of fixed tension in the $(d+2)$-dimensional theory has $E \propto 1/r^{z-2\theta/d}$. Probe scalar fields will be discussed in \S \ref{sec:scalars}.

Before proceeding, it is important to point out that the metric (\ref{eq:dsIR1}) will only give a good description of the dual theory in a certain range of $r$, and there could be important corrections for $r \to 0$ or very large $r$. Outside the range with hyperscaling violation, 
but assuming spatial and time translation symmetries and spatial rotation invariance,
the metric will be of the more general form
\be\label{eq:dsgeneral}
ds_{d+2}^2= e^{2A(r)} \left(-e^{2B(r)} dt^2 + dr^2 +dx_i^2\right)\,.
\ee
An important situation corresponds to a field theory that starts from a UV fixed point and then develops a scaling violation exponent $\theta$ at long-distances. This means that the gravity side warp factor $e^{2A} \to R^2/r^2$ for $r \to 0$ (with $R$ the AdS radius) and that below a cross-over scale $r_F$ it behaves as in (\ref{eq:dsIR1}). This scale then appears in the metric as an overall factor $ds^2 \propto R^2/r_F^{2\theta/d}$, and is responsible for restoring the canonical dimensions in the presence of hyperscaling violation.\footnote{For instance, in models with a Fermi surface, $r_F$ is set by the Fermi momentum~\cite{Ogawa:2011bz}.} Finally, at scales $r \gg r_F$ the theory may flow to some other fixed point, develop a mass gap etc., and (\ref{eq:dsIR1}) would again no longer be valid. String theory examples that exhibit these flows will be presented in \S \ref{sec:string}. For now we will simply ignore these corrections and focus on the form (\ref{eq:dsIR1}), keeping in mind that it may be valid only in a certain window of energies. We follow an `effective' approach where the dual theory is taken to live at finite $r$ of order $r_F$.

In order to understand the metric properties of this class of spacetimes, notice that (\ref{eq:dsIR1}) is conformally equivalent to a Lifshitz geometry, as can be seen after a Weyl rescaling
$g_{\mu \nu } \to \t g_{\mu\nu}=\Omega^2 g_{\mu\nu}$, with $\Omega= r^{-\theta/d}$. (The scale-invariant limit is $\theta=0$, which reduces to a Lifshitz solution.) Since a Lifshitz metric has constant curvature, the scalar curvature associated to (\ref{eq:dsIR1}) acquires $r$-dependent terms controlled by the derivative of the Weyl factor $\Omega$. 

The Appendix contains the Ricci and Einstein tensors for the general class of metrics (\ref{eq:dsgeneral}). In particular, the Ricci tensor for the metrics (\ref{eq:dsIR1}) is given by
\bea\label{eq:Rmunu1}
R_{tt}&=&\frac{(d+z-\theta ) (d z-\theta )}{d r^{2z}} \nonumber\\
R_{rr}&=&\frac{-d \left(d+z^2\right)+(d+z) \theta }{d r^2}\nonumber\\
R_{ij}&=&-\delta_{ij}\frac{(d-\theta ) (d+z-\theta )}{d r^2}\,.
\eea
The scalar curvature is then $R \propto r^{-2 \theta /d}$, which becomes constant for $\theta=0$ as expected.

\subsection{Constraints from the null energy condition}\label{subsec:NEC}

What types of constraints should we impose on (\ref{eq:dsIR1}) in order to get a physically sensible dual field theory? Quite generally, from the gravity side we should demand at least that the null energy condition (NEC) be satisfied. That is, we impose 
\be
T_{\mu\nu}N^\mu N^\nu\geq0
\ee
on the Einstein equations, where $N^\mu N_\mu = 0$. Since $G_{\mu\nu}=T_{\mu\nu}$ on-shell, from (\ref{eq:Rmunu1}) the constraints from the NEC become\footnote{For the general metric (\ref{eq:dsgeneral}), the two independent null vectors are 
\be
N^t=\;,\;N^r=e^{-A}\cos\varphi\;,\;N^i=e^{-A}\sin\varphi,\nonumber
\ee
where $\varphi=0$ or $\pi/2$.
}
\bea\label{eq:NEC1}
&&(d-\theta)\left(d(z-1)-\theta\right)\geq 0\nonumber\\ 
&&(z-1)(d+z-\theta)\geq 0\,.
\eea

The constraints (\ref{eq:NEC1}) have important consequences for the allowed values of $(z, \theta)$ that admit a consistent gravity dual. First, in a Lorentz invariant theory, $z=1$ and then the first inequality implies that $\theta \le 0$ or $\theta \ge d$. Both ranges will be realized in the string theory constructions of \S \ref{sec:string}. On the other hand, for a scale invariant theory ($\theta=0$), we recover the known result $z \ge 1$.

Theories with $\theta=d-1$ are of interest since they give holographic realizations of theories with several of the properties of Fermi surfaces~\cite{Ogawa:2011bz,Huijse:2011ef}. The NEC then requires that the dynamical exponent satisfies
\be
z \ge 2 - 1/d\,,
\ee
in order to have a consistent gravity description. More generally, in \S \ref{sec:entanglement} we will find that for systems with
\be
d-1 \le \theta \le d\,,
\ee
the entanglement entropy exhibits new violations of the area law. These cases can be realized for a dynamical exponent that satisfies $z \ge 1 - \theta /d$. The limit $\theta=d$ will correspond to an extensive violation of the entanglement entropy, and requires $z \ge 1$ or $z \le 0$.

Notice that in theories with hyperscaling violation the NEC can be satisfied for $z<1$, while this range of dynamical exponents is forbidden if $\theta=0$ \cite{Koroteev}. In particular, $\{z<0, \theta>d\}$ gives a consistent solution to (\ref{eq:NEC1}), as well as $\{0<z<1, \theta \ge d+z\}$. Notice that, just based on the NEC, the range $\theta > d$ is allowed. Clearly more information is needed to determine whether the above choices lead to physically consistent theories -- in particular we will argue below that $\theta>d$ leads to instabilities in the gravity side. In what follows we continue this analysis using holographic techniques to calculate correlation functions, entanglement entropy and thermal effects. It would also be interesting to derive conditions for the existence of these theories directly in the field theory side.

\subsection{Massive propagators}\label{subsec:propagators1}

The next step is to calculate two-point functions $\langle \mc O(x) \mc O(x') \rangle$, where $\mc O$ is some operator in the dual theory. We will consider an operator that can be described by a scalar field in the bulk. The simplest correlation functions correspond to massive propagators in the bulk, since in the semiclassical approximation this is given in terms of the geodesic distance traveled by a particle of mass $m$. In AdS/CFT, massive bulk propagators give power-law CFT Green's functions because of the $r$-dependent warp factor. The geodesic distance is minimized by moving into the bulk, and this turns an exponential into a power-law; see e.g.~\cite{Susskind:1998dq}.

Let us now calculate correlation functions in the semiclassical approximation, for the class of metric (\ref{eq:dsIR1}), in the range $\theta \leq d$. The full correlator away from the semiclassical limit will be studied in \S \ref{sec:scalars}.
The particle geodesic describing the semiclassical trajectory is obtained by extremizing the action
\be\label{eq:Sparticle}
S = - m \int d \lambda \, r^{-(d-\theta)/d}\sqrt{-r^{-2(z-1)} \dot t^2 + \dot r^2 + \dot x^2}
\ee
where $\lambda$ is the worldline coordinate and a `dot' denotes a derivative with respect to $\lambda$. The propagator between $x = (t,x_i)$ and $x'=(t',x_i')$ on a fixed $r = \epsilon$ surface is then given by
\be\label{eq:prop1}
G_\epsilon (x,x') \sim e^{S(x,x')}
\ee
with the conditions $(x(0)=x, r(0)= \epsilon)$ and $(x(1)=x', r(1)= \epsilon)$. Because of time and space translation invariance, the propagator only depends on $\Delta t \equiv t'-t$ and $\Delta x_i= x_i'- x_i$. Here the cutoff $\epsilon \sim r_F$, but otherwise it is left unspecified.

Consider first the case of spacelike propagation, with $\Delta t=0$. Choosing $\lambda =r$ gives
\be\label{eq:Sparticle1}
S = - m \int dr \, r^{-(d-\theta)/d}\sqrt{1 + (dx/dr)^2}\,.
\ee
Since the momentum conjugate to $x$ is conserved, the equation of motion is integrated to
\be\label{eq:dx}
\frac{dx}{dr}= \frac{(r/r_t)^{(d-\theta)/d}}{\sqrt{1-(r/r_t)^{2(d-\theta)/d}}}\,.
\ee
Here $r_t$ is the turning point for the geodesic, $dr/dx|_{r=r_t}=0$. It is related to $\Delta x$ by integrating (\ref{eq:dx}):
\be
\frac{|\Delta x|}{2} = \frac{\sqrt{\pi}\,\Gamma\left(\frac{2d-\theta}{2(d-\theta)} \right)}{\Gamma \left(\frac{d}{2(d-\theta)} \right)} r_t\,.
\ee

Plugging (\ref{eq:dx}) into (\ref{eq:Sparticle}), we obtain the geodesic distance
\be
S = \frac{2 d}{\theta}\,m\, \epsilon^{\theta/d}  -\frac{d}{\theta} c_{\theta,\,d}\,m\,|\Delta x|^{\theta/d}\,,
\ee
where we have neglected higher powers of $\epsilon$ and defined
\be
c_{\theta,\,d} \equiv \left(\frac{2\sqrt{\pi}\,\Gamma\left(\frac{2d-\theta}{2(d-\theta)} \right)}{\Gamma \left(\frac{d}{2(d-\theta)} \right)} \right)^{(d-\theta)/d}\,.
\ee
Thus, the propagator in the semiclassical approximation becomes
\be\label{eq:prop2}
G_\epsilon(\Delta x) \sim  \exp\left[\frac{2 d}{\theta}\,m\, \epsilon^{\theta/d}\right] \exp\left[-\frac{d}{\theta} c_{\theta,\,d}\,m\,|\Delta x|^{\theta/d}\right]\,.
\ee
The approximation holds in the regime
\be\label{eq:WKB}
m\,|\Delta x|^{\theta/d} \gg 1\,,
\ee 
in units of the cross-over scale $r_F$.

As a check, in the scale-invariant limit $\theta= 0$, the integrals for $r_t$ and $S$ give logarithms instead of powers, and we should replace $\epsilon^{\theta/d}/(\theta/d) \to \log \epsilon$ (and similarly for $r_t$). Then
\be
G_\epsilon(\Delta x) \sim \exp \left[m \log \frac{\epsilon}{|\Delta x|} \right] \sim \frac{\epsilon^m}{|\Delta x|^m}\,,
\ee
reproducing the expected CFT power-law behavior.

The correlator (\ref{eq:prop2}) reveals interesting properties about the dual field theory with hyperscaling violation in the WKB regime. First, it has an exponential (rather than power-law) dependence on $|\Delta x|$, showing that the dual theory is massive, with a nontrivial RG evolution -- at least for operators dual to massive scalars. However, the usual weakly coupled decay $\sim \exp(-|\Delta x|)$ is now replaced by a nontrivial $\theta$-dependent exponent. For $\theta>0$ the propagator decays exponentially at large distances with an exponent $|\Delta x|^{\theta/d}$. For $\theta<0$ the propagator appears to approach a constant value at large distances, but it is outside the regime \eqref{eq:WKB} where the semiclassical approximation is valid.\footnote{In the string theory realizations of \S \ref{sec:string}, the theories with $\theta <0$ eventually exit the regime with hyperscaling violation, modifying this propagator behavior.} 

Anticipating our results of \S \ref{sec:scalars}, we point out that away from the semiclassical limit (or more generally for operators dual to massless scalars) the propagators do exhibit power-law behavior, with a power that includes a shift by $\theta$.

So far our calculations have been for a spacelike geodesic; similar computations lead to the correlator for a timelike path, now with nontrivial $z$ dependence. Working in Euclidean time, $\tau=it$, the value of the action becomes,
\be
S={2d\over \theta}m\epsilon^{\theta/d}-m\left ({dz\over \theta}\right )^{\theta/dz}\tilde{c}_{d,\theta,z}|\Delta\tau|^{\theta/dz},
\ee
where
\be
\tilde{c}_{d,\theta,z}=\left (\frac{(z-\theta/d)\Gamma\left (\frac{z}{2(z-\theta/d)}\right )}{\sqrt{\pi}\Gamma\left (\frac{\theta/d}{2(z-\theta/d)}\right )}\right )^{\theta/dz-1}.
\ee
This is valid in the range $0 < \theta/d < z.$ We see that for the case $z=1$, this reduces to the solution for the spacelike geodesic. The propagator for a timelike path is
\be
G_\epsilon(\Delta\tau) \sim  \exp\left[\frac{2 d}{\theta}\,m\, \epsilon^{\theta/d}\right] \exp\left[-\left (\frac{dz}{\theta}\right )^{\theta/dz} \tilde{c}_{d,\theta,z}\,m\,|\Delta \tau|^{\theta/dz}\right]
\ee
in the regime where
\be
m|\Delta\tau|^{\theta/dz}\gg 1.
\ee

The propagator for an arbitrary geodesic is in general a function of  both $|\Delta x|$ and $|\Delta\tau |$. Now that we have discussed the two specific extremes, we briefly discuss the general solution. The differential equations cannot be solved analytically for arbitrary $d,\theta,z$, but can in principle be solved numerically for specific values of the critical exponents. We outline this procedure in appendix \ref{sec:geodesics}.

\vskip 2mm

\subsection{Holographic energy-momentum tensor}\label{subsec:energyT}

Another important object that characterizes the dual QFT is the expectation value of the energy-momentum tensor. It contains information about the number of degrees of freedom (e.g. the central charge in a 2d theory) and other conformal anomalies. In order to calculate the stress tensor, we need a method that can be applied locally to a radial slice, and which does not assume an asymptotic AdS structure -- after all, the metric (\ref{eq:dsIR1}) may give a good description of the QFT only in an open range of radial scales. An adequate method for this case is to compute the Brown-York stress tensor~\cite{Brown:1992br} on a radial slice, and identify it with the expectation value of the energy-momentum tensor in the dual theory~\cite{Balasubramanian:1999re,Horowitz:1998ha}.\footnote{This has been recently applied to de Sitter and FRW solutions, which naturally have a radial cutoff,  in~\cite{Dong:2011uf}.}

The basic idea is as follows. Consider a hypersurface of constant $r=r_c$ and let $n_\mu$ be the unit normal vector to this timelike surface. For us, $r_c \sim \mc O (r_F)$. The induced metric is
\begin{equation}
\gamma_{\mu\nu}=g_{\mu\nu}-n_{\mu}n_{\nu}\,,
\end{equation}
and the extrinsic curvature is given by
\begin{equation}
K_{\mu\nu}=-{\gamma_{\mu}}^{\rho}\nabla _{\rho}n_{\nu}\,.
\end{equation}
Since $r_c$ will be taken to be finite (for instance, of order of the cross-over scale), counterterms and regularization issues will be ignored. Then 
the quasilocal stress tensor~\cite{Brown:1992br} is
\begin{equation}
\tau_{\mu\nu}= K_{\mu\nu}-\gamma_{\mu\nu} K^\rho_{\,\rho}\,,
\end{equation}
ignoring a dimensionful constant.

The AdS/CFT correspondence relates the expectation value of the stress tensor $\langle T_{\mu\nu}\rangle$ in the dual  theory to the limit of the quasilocal stress tensor $\tau_{\mu\nu}$ as $r_c \to 0$ (the boundary):
\begin{equation}\label{Tcfres}
\sqrt{-h}h^{\mu \rho}\langle T_{\rho \nu}\rangle=\lim_{r_c\to 0}\sqrt{-\gamma}\gamma^{\mu\rho}\tau_{\rho \nu},
\end{equation}
where $h_{\mu\nu}$ is the background QFT metric, which is related to $\gamma_{\mu\nu}$ by a conformal transformation. Our proposal is to extend this relation to metrics of the form (\ref{eq:dsIR1}) at finite $r$, and use $\tau_{\mu\nu}$ to determine $\langle T_{\mu \nu}\rangle$.

A radial slice at $r=r_c$ has an induced metric
\be
\gamma_{\mu\nu} dx^\mu dx^\nu =r_c^{-2(d-\theta)/d} \left(-r_c^{-2(z-1)} dt^2 + dx_i^2 \right)\,,
\ee
and the background QFT metric is given by $h_{\mu\nu}= r_c^{2(d-\theta)/d} \gamma_{\mu\nu}$. Let us first consider for simplicity the Lorentz invariant case. A metric of the form
\be
ds^2 = dw^2 + h(w)^2 \eta_{\mu\nu} dx^\mu dx^\mu 
\ee
has extrinsic curvature $K_{\mu\nu}= - h(w) \partial_w h(w)\,\eta_{\mu\nu}$
at constant $w$. Applying this to our case and using (\ref{Tcfres}) obtains
\be\label{eq:Tmunu}
\langle T_{\mu\nu} \rangle = - \frac{d-\theta}{r_c^{d+1 - \theta}} h_{\mu\nu}\,.
\ee
In the more general case of $z\neq1$, we break Lorentz invariance and the nontrivial components of the energy-momentum tensor can be determined as
\be
\langle T_{00}\rangle=-\frac{d-\theta}{r_c^{d+1-\theta}}h_{00}\,,\quad
\langle T_{ij}\rangle=-\frac{zd-\theta}{r_c^{d+1-\theta}}h_{ij}\,.
\ee

In a $(d+1)$-dimensional CFT, the energy-momentum tensor is an operator of conformal weight $d+1$, so we expect $\langle T_{\mu\nu} \rangle \sim h_{\mu\nu}/r_c^{d+1}$. More precisely, a nonvanishing one-point function is obtained by placing the CFT on a curved background of constant curvature, and then $\langle T_{\mu\nu} \rangle \sim h_{\mu\nu}/R^{d+1}$ with $R$ the curvature radius. Obtaining this from the gravity side requires adding counterterms and taking the limit $r_c \to 0$. Here we are working at finite $r_c$ and we ignore these subtraction terms, since in general the metric with hyperscaling violation is not valid near the boundary.

Hyperscaling violation has then the effect of shifting the energy-momentum tensor one-point function to $h_{\mu\nu}/r_c^{d+1 - \theta}$. A similar result will be obtained in correlators of marginal operators below. From this point of view, a possible interpretation is that $\theta$ reflects a nonzero scaling dimension for the vacuum.\footnote{We thank S. Shenker for interesting remarks along these lines.} However, the effects of $\theta$ in the field theory are probably more complicated than just a universal shift in the vacuum. We will return to these points in \S \ref{sec:scalars}.

\section{Dynamics of scalar operators}\label{sec:scalars}

Having understood the basic properties of holographic theories with hyperscaling violation, in this section we study in detail operators that are described by bulk scalar fields with action
\be
S = -\frac{1}{2}\, \int d^{d+2}x\,\sqrt{g}\, \left(g^{\mu\nu} \partial_\mu \phi \partial_\nu \phi + m^2 \phi^2 \right)\,.
\ee
In particular, we will analyze two-point functions valid for arbitrary (not necessarily large) mass $m$, where the WKB approximation of \S \ref{subsec:propagators1} is not applicable.

\subsection{Scalar field solution}\label{subsec:scalarsol}

The equation of motion for a scalar field with mass $m$ in the background (\ref{eq:dsIR1}) is
\be\label{eq:scaeom}
\left(\partial_r^2-\frac{d-\theta+z-1}{r}\partial_r+\partial_i^2-r^{2(z-1)}\partial_t^2-\frac{m^2}{r^{2(d-\theta)/d}}\right)\phi=0\,.
\ee
Let us first consider the behavior of $\phi$ at small $r$. Starting from an ansatz $\phi\sim r^\alpha$, we find that the $\partial_i$, $\partial_t$, and $m^2$ terms are all subdominant at small $r$ if $z>0$ and $\theta>0$. In this case, we can solve the equation at leading order in $r$, which gives $\alpha=0$ or $d-\theta+z$. This means that when we impose the incoming boundary condition at $r=\infty$ (or in the Euclidean picture, the regularity condition), the full solution has the following expansion around $r=0$:
\be\label{eq:scasol}
\phi=1+\cdots+G(\vec k,\omega)r^{d-\theta+z}(1+\cdots)\,,
\ee
where we have Fourier transformed in the $t$ and $\vec x$ directions, and $\cdots$ refers to higher order terms in $r$.

The behavior in \eqref{eq:scasol} should be contrasted to the case of $\theta=0$, in which the mass term becomes one of the leading contributions, and we are back to the Lifshitz or AdS case of $\alpha=\frac{d+z}{2}\pm\sqrt{\left(\frac{d+z}{2}\right)^2+m^2}$.

The momentum-space two-point function on the boundary for the operator dual to $\phi$ is given by the coefficient function $G(\vec k,\omega)$ in the expansion above \cite{Sandip}. We will analyze its behavior in the next few subsections, while solving it exactly in a few special cases.

\subsection{Massless case}\label{subsec:massless}
For simplicity we will consider the $z=1$ case where we recover Lorentz invariance. The equation of motion \eqref{eq:scaeom} becomes exactly solvable for a massless scalar:
\be
\left(\partial_r^2-\frac{d-\theta}{r}\partial_r-k^2\right)\phi=0\,,
\ee
with $k=(\omega,\vec k)$. The solution that satisfies the correct boundary condition at $r=\infty$ is
\be
\phi=(kr)^{(d-\theta+1)/2}K_{(d-\theta+1)/2}(kr)\ .
\ee
Note that we have normalized the solution at the boundary according to \eqref{eq:scasol}, up to a numerical factor that does not depend on $k$. Expanding the modified Bessel function, we find (again up to a $k$-independent constant)
\be\label{eq:masslk}
G(k)\sim k^{d-\theta+1} .
\ee
Fourier transforming back to position space, we find the two-point function to be
\be\label{eq:masslx}
\langle\mc O(x)\mc O(x')\rangle=\int\frac{d^{d+1}k}{(2\pi)^{d+1}}G(k)e^{ik\cdot(x-x')}\sim\frac{1}{|x-x'|^{2(d+1)-\theta}}\,.
\ee
Here $\mc O$ is a marginal operator dual to the massless $\phi$ in the bulk. We find that the dimension of this marginal operator is shifted by $\theta$.

\subsection{Massive case: a scaling argument}\label{subsec:massive}
In the more general case where the mass is nonzero, we cannot solve the scalar equation of motion
\be\label{eq:sleom}
\left(\partial_r^2-\frac{d-\theta}{r}\partial_r-k^2-\frac{m^2}{r^{2(d-\theta)/d}}\right)\phi=0
\ee
in closed form (except for special values of $\theta$ which we will discuss in the next subsection). However, we note a scaling symmetry in the equation
\be
r\to\lambda r\,,\quad
k\to k/\lambda\,,\quad
m\to m/\lambda^{\theta/d}\,,
\ee
under which the coefficient function $G(k)$ should transform as
\be
G(k;m)=\lambda^{d-\theta+1}G(k/\lambda;m/\lambda^{\theta/d})
\ee
in order to keep \eqref{eq:scasol} invariant.

We immediately observe that in the massless case $G(k)\sim k^{d-\theta+1}$, by setting $\lambda=k$. This agrees with our results in the previous subsection.

We also find that for positive $\theta$, the mass term become unimportant at short distances, and the UV behavior of the massive two-point function is given by the massless results (\ref{eq:masslk}, \ref{eq:masslx}). The long-distance behavior of the massive two-point function is given by the WKB approximation of \S\ref{subsec:propagators1}. We will verify these statements in a few exactly solvable cases in the next subsection.

When $\theta$ is negative, the mass term becomes unimportant at long distances, and the IR behavior of the massive two-point function is given by the massless results.

We have restricted ourselves to the $z=1$ case here, but our results apply more generally for $z\neq1$ as well. In that case the scaling symmetry is
\be
r\to\lambda r\,,\quad
\vec k\to\vec k/\lambda\,,\quad
\omega\to\omega/\lambda^z\,,\quad
m\to m/\lambda^{\theta/d}\,,
\ee
and the momentum-space two-point function transforms as
\be
G(\vec k,\omega;m)=\lambda^{d-\theta+z}G(\vec k/\lambda,\omega/\lambda^z;m/\lambda^{\theta/d})\,.
\ee
Fourier transforming back to position space, we have
\be
G(\Delta\vec x,\Delta t;m)=\lambda^{2(d+z)-\theta}G(\lambda\Delta\vec x,\lambda^z\Delta t;m/\lambda^{\theta/d})\,.
\ee
The equal-time two-point function in the massless case is therefore given by
\be
G(\Delta\vec x)\sim\frac{1}{|\Delta\vec x|^{2(d+z)-\theta}}\,.
\ee

\subsection{Some special cases}

The equation of motion \eqref{eq:sleom} simplifies and becomes solvable in some special cases, which we now discuss.

\subsubsection{The case $\theta=d$}

 The solution that satisfies the correct normalization and boundary condition is
\be
\phi=\exp\left(-\sqrt{k^2+m^2}\,r\right)\,.
\ee
The two-point function in momentum space is therefore
\be
G(k)=\sqrt{k^2+m^2}\,.
\ee
At short distance, the two-point function is dominated by the large-$k$ behavior $G(k)\sim k$, and we have
\be
\langle\mc O(x)\mc O(x')\rangle\sim\frac{1}{|x-x'|^{d+2}}\quad
\text{for small }|x-x'|\,,
\ee
which agrees with \eqref{eq:masslx} for $\theta=d$.

At long distance, the two-point function in position space
\be
\langle\mc O(x)\mc O(x')\rangle=\int\frac{d^{d+1}k}{(2\pi)^{d+1}}\sqrt{k^2+m^2}e^{ik\cdot(x-x')}
\ee
can be shown to decay as $e^{-m|x-x'|}$ by the saddle point approximation. This agrees with \eqref{eq:prop2} for $\theta=d$.

\subsubsection{The case $\theta=d/2$}

The equation of motion \eqref{eq:sleom} is exactly solvable for $\theta=d/2$ in terms of the confluent hypergeometric function. A special case of this kind is $d=2$ and $\theta=1$, which is a candidate holographic realization of a Fermi surface in $2+1$ dimensions 
\cite{Huijse:2011ef}.

For $\theta=d/2$ with general $d$, the solution with the correct normalization and boundary condition is
\begin{align}
\phi&=\frac{\Gamma(1+\frac{d}{4}+\frac{m^2}{2k})}{\Gamma(1+\frac{d}{2})}\,e^{-kr}(2kr)^{1+\frac{d}{2}}U\left(-\frac{d}{4}+\frac{m^2}{2k}\,,\,-\frac{d}{2}\,,\,2kr\right)\\
&=1+\cdots+(2kr)^{1+\frac{d}{2}}\frac{\Gamma(-1-\frac{d}{2})\Gamma(1+\frac{d}{4}+\frac{m^2}{2k})}{\Gamma(1+\frac{d}{2})\Gamma(-\frac{d}{4}+\frac{m^2}{2k})}+\cdots\,,
\end{align}
from which we read off the two-point function in momentum space:
\be\label{eq:Gkd2}
G(k)=(2k)^{1+\frac{d}{2}}\frac{\Gamma(-1-\frac{d}{2})\Gamma(1+\frac{d}{4}+\frac{m^2}{2k})}{\Gamma(1+\frac{d}{2})\Gamma(-\frac{d}{4}+\frac{m^2}{2k})}\,.
\ee
Again, the short-distance behavior of the two-point function is given by $G(k)\sim k^{1+d/2}$ at large $k$, and agrees with \eqref{eq:masslx} for $\theta=d/2$.

When $d$ is even, the first gamma function in the numerator of \eqref{eq:Gkd2} diverges and indicates the appearance of logarithmic terms in $k$. The two-point function in momentum space becomes
\be
G(k)\sim k^2\log k
\ee
at large $k$, which gives us $G(\Delta x)\sim 1/|\Delta x|^3$.

\section{Entanglement entropy}\label{sec:entanglement}

In this section we evaluate the entanglement entropy for systems with hyperscaling violation, according to the holographic proposal of Ryu and Takayanagi~\cite{Ryu:2006bv}. Our main result is the entropy formula (\ref{eq:entgltheta}) for theories with arbitrary $(z, \theta)$. We will use this to probe various properties of these theories, including ground state degeneracies and the appearance of Fermi surfaces. Our study will reveal novel phases for $d-1 \le \theta \le d$, which feature violations of the area law that interpolate between logarithmic and linear behaviors.

A natural question in a holographic study of entanglement entropy is which system extremizes the entanglement entropy over a given class of metrics.
This question is motivated in the following sense: one measure of strong correlation is ground-state entanglement, and it is 
well known that some of the most interesting systems (Fermi liquids, non-Fermi liquids) have entanglement which scales more
quickly with system size than `typical' systems.  It is therefore worthwhile to ask, does holography indicate new phases (dual
to new bulk metrics) with equally large or larger anomalous ground-state entanglement?
This was one of our original motivations in this analysis. In \S \ref{subsec:extreme} we address this question for metrics with hyperscaling violation, finding that $\theta = d-1$ is the only local extremum. This implies that systems with a Fermi surface minimize the entanglement.

\subsection{General analysis}\label{subsec:general}

Before computing the entanglement entropy in systems with hyperscaling violation, let us discuss the more general class of metrics
\be\label{eq:dsgeneral2}
ds_{d+2}^2= e^{2A(r)} \left(-e^{2B(r)} dt^2 + dr^2 +dx_i^2\right)\,.
\ee
We will first calculate the entanglement entropy across a strip. This is the simplest case to analyze. Then it will be argued that the same behavior is found for general entanglement regions when their diameter is large.

Let us then begin by computing the entanglement entropy for a strip
\be
-l \le x_1 \le l\;,\;0 \le x_i \le L\;,\;i = 2,\,\ldots, d
\ee
in the limit $l \ll L$. We focus on the case of $\theta\le d$ so the strip is located on a UV slice at $r=\epsilon$. The profile of the surface in the bulk is $r=r(x_1)$, and its area is
\be\label{eq:area1}
\mc A = L^{d-1} \int_0^{r_t} e^{dA(r) }\sqrt{1+ \left(\frac{dx_1}{dr} \right)^2}\,.
\ee
We have inverted $x_1=x_1(r)$ to make the conserved momentum manifest, and the turning point $r_t$ corresponds to $dr/dx_1|_{r_t}=0$. To obtain the entanglement entropy we need to extremize $\mc A$ and evaluate it on the dominant trajectory.

The calculation follows the same steps as those of \S \ref{subsec:propagators1} for the particle geodesic. Extremizing $\mc A$ and using the conserved momentum obtains
\be\label{eq:trajectory}
\frac{dx_1}{dr} = \frac{e^{-d \left(A(r) - A(r_t) \right)}}{\sqrt{1-e^{-2d \left(A(r) - A(r_t) \right)}}}\,.
\ee
The turning point is thus fixed in terms of the length $l$ by the integral of this expression,
\be\label{eq:l}
l = \int_0^{r_t} dr \, \frac{e^{-d \left(A(r) - A(r_t) \right)}}{\sqrt{1-e^{-2d \left(A(r) - A(r_t) \right)}}}\,.
\ee
Finally, replacing (\ref{eq:trajectory}) into (\ref{eq:area1}) obtains the formula for the area
\be\label{eq:area2}
{\mc A} = L^{d-1} \int_\epsilon^{r_t} dr\,\frac{e^{d A(r)}}{\sqrt{1-e^{-2d \left(A(r) - A(r_t) \right)}}}\,.
\ee
The entanglement entropy for a strip in the general metric (\ref{eq:dsgeneral}) is thus 
\be
\mc S = \frac{M_{Pl}^d}{4} \mc A
\ee
with $M_{Pl}$ the $(d+2)$ -dimensional Planck constant. 

\subsubsection{General entanglement regions}

While most of our analysis will be carried out explicitly for the simplest case of a strip, our conclusions will also apply to general entanglement surfaces. We will now establish this, by showing that the entanglement entropy for a general surface is given approximately by (\ref{eq:area2}) both near the boundary and at long distances.

Consider a general surface, parametrized by
\be
x_d = \sigma(x_i)\;,\;i =1,\ldots, d-1
\ee
at $r=0$. The surface that extremizes the area will then be of the form
\be
x_d = \Sigma(x_i, r)\;,\;\Sigma(x_i,0) = \sigma(x_i)\,.
\ee
The pullback of the bulk metric onto $\Sigma$ gives
\be
ds_\Sigma^2 =e^{2A}\left(\left[ 1 + (\partial_r \Sigma)^2\right] dr^2 + 2  \partial_r \Sigma \partial_i \Sigma\, dr dx_i +  \left(\delta_{ij}+ \partial_i \Sigma \partial_j \Sigma \right) dx^i dx^j\right)
\ee
and the area reads
\be\label{eq:areaS}
\mc A = \int d^{d-1} x \,dr \,e^{dA(r)} \sqrt{1+ (\partial_i \Sigma)^2+(\partial_r \Sigma)^2}\,.
\ee
The equation of motion implies the existence of a conserved current $J_M$ with components
\bea\label{eq:JM}
J_r &=& e^{dA(r)}\,\frac{\partial_r \Sigma}{\sqrt{1+ (\partial_i \Sigma)^2+ f(r) (\partial_r \Sigma)^2}} \nonumber\\
J_i&=&e^{dA(r)}\,\frac{\partial_i \Sigma}{\sqrt{1+ (\partial_i \Sigma)^2+ (\partial_r \Sigma)^2}}\,.
\eea
Integrating $\partial_M J^M=0$ over $x_i$, we read off the conserved charge
\be
Q_r = \int d^{d-1} x\,e^{dA(r)}\, \frac{\partial_r \Sigma}{\sqrt{1+ (\partial_i \Sigma)^2+  (\partial_r \Sigma)^2}}\,,
\ee
which generalizes the result for a strip (\ref{eq:trajectory}) to an arbitrary shape.

We will now show that (\ref{eq:areaS}) reduces to the case of a strip (\ref{eq:area2}) both near the boundary $r= \epsilon$ and near a `turning point' $\partial_r \Sigma \to \infty$. First, for $r \to 0$, $\Sigma(x_i,r)$ may be expanded as\footnote{A similar analysis near the boundary for a specific metric appeared recently in~\cite{Huijse:2011ef}.}
\be
\Sigma(x_i,r) = \sigma(x_i) + r^\lambda \sigma_1(x_i) + \ldots\,.
\ee
The equation of motion for $\sigma_1$ then requires $\lambda=1$.
Then the UV part of the area is of the form
\be
\mc A_{UV} \approx \left(\int d^{d-1} x\,\sqrt{1+ (\partial_i \sigma)^2} \right) \int_\epsilon dr \,e^{d A(r)}\,,
\ee
which indeed agrees with the result for the strip (\ref{eq:area2}). 

Now we want to show that for regions of large area (or diameter), the long-distance part of the entanglement entropy also coincides with (\ref{eq:area2}). Intuitively, when the size of the system is large most of the surface is deep inside the bulk, and the scaling of the entropy can be approximated by the behavior in the vinicity of a turning point $r=r_t$ with $\partial_r \Sigma \to \infty$.

In more detail, we require that locally around $r_t$, $\partial_i \Sigma$ is smooth and that the combination
$$
e^{dA(r)}  (r-r_t)^{1/2} \to 0\;\;\textrm{as}\;\;r\to r_t\,.
$$
This guarantees that $J_i \to 0$ as $r \to r_t$ (see (\ref{eq:JM}) for a definition of $J_M$) and hence the current conservation equation implies that $J_r \approx \textrm{const}$ near $r=r_t$. In this approximation,
\be
\partial_r \Sigma \approx \sqrt{1+(\partial_i \Sigma)^2}\, 
 \frac{e^{-d\left(A(r)-A(r_t)\right)}}{\sqrt{1-e^{-2d\left(A(r)-A(r_t)\right)}}}\,.
\ee
This agrees with the behavior (\ref{eq:trajectory}), so the entropy from the IR region also agrees with that for a strip,
\be
\mc A_\textrm{IR} \approx  \left(\int d^{d-1} x\,\sqrt{1+ (\partial_i \Sigma)^2} \right) \int^{r_t} dr\,\frac{e^{d A(r)}}{\sqrt{1-e^{-2d \left(A(r) - A(r_t) \right)}}}\,.
\ee

Given these results, in what follows our calculations will be done explicitly for a strip, keeping in mind that our conclusions are valid for more general entanglement regions that satisfy the above criterion.

\subsubsection{Using trial surfaces}

Let us also briefly mention that the basic properties of the entanglement entropy can be understood by considering a simple trial surface in the bulk (e.g. a cylinder) and requiring that it is a stationary point. This method also applies to a general entanglement region.

Consider a general entanglement region $\Sigma$ defined on a slice $r=\epsilon$; denote its volume $|\Sigma|$ and surface area by $|\partial \Sigma|$. We now approximate the bulk surface used in the holographic calculation of the entanglement entropy by a cylinder with boundary $\partial \Sigma$, that extends from $r=\epsilon$ to $r=r_t$. The value $r_t$ is chosen such that it extremizes the entanglement entropy.\footnote{Of course, this approach is well-known in applications of the method of variations. See e.g.~\cite{Barbon} for an earlier work in the context of entanglement entropy.} Starting from the general metric (\ref{eq:dsgeneral2}), the total area of this bulk cylinder is then
\be\label{eq:trialA}
\mc A = |\partial \Sigma| \int_\epsilon^{r_t}\,dr\,e^{dA(r)} +|\Sigma| e^{dA(r_t)}\,.
\ee
Requiring that $r_t$ is a stationary point gives
\be\label{eq:stationary}
A'(r_t) = - \frac{1}{d} \frac{|\partial \Sigma|}{|\Sigma|} \equiv - \frac{1}{dl}
\ee
where we introduced the perimeter $l \equiv |\Sigma|/|\partial \Sigma|$.

Given a concrete warp factor, (\ref{eq:stationary}) determines the value of $r_t$, and then (\ref{eq:trialA}) gives the approximation of the entanglement entropy by trial cylinders. The stationary point is a minimum or a maximum depending on the sign of $A''(r)$:
\be\label{eq:trialmass}
\frac{\delta^2 \mc A}{\delta r_t^2} =|\Sigma| d \,e^{dA(r_t)} A''(r_t)\,,
\ee
evaluated at the critical point.
In examples below we will find that this procedure gives a good qualitative understanding of the entanglement entropy.

\subsection{Entanglement entropy with hyperscaling violation}\label{subsec:Shyperscaling}

Now we are ready to evaluate the entanglement entropy for metrics with hyperscaling violation. It is useful to start by recalling the scale invariant $\theta=0$ case. The bulk metric corresponds to $e^{2A} = R^2/r^2$ in the ansatz (\ref{eq:dsgeneral2}), with $R$ the AdS radius.
Plugging this into (\ref{eq:l}) and (\ref{eq:area2}) obtains
\be\label{eq:entaglAdS}
\mc S = \frac{(R M_{Pl})^d}{4(d-1)} \left[\left(\frac{L}{\epsilon}\right)^{d-1}- \left(\sqrt{\pi} \,\frac{\Gamma\left(\frac{1+d}{2d} \right)}{\Gamma\left(\frac{1}{2d} \right)}\right)^d \left(\frac{L}{l} \right)^{d-1} \right]\,,
\ee
up to higher powers of $\epsilon$. For $d=3$, this reproduces the entanglement entropy for $\mc N=4$ SYM, after relating the 5d Einstein frame quantities to their 10d counterparts.

The hyperscaling violation exponent modifies the warp factor to $e^{dA(r)}=r^{\theta -d}$ (ignoring for now the cross-over scale). Now (\ref{eq:area2}) gives
\be\label{eq:A}
\mc A =L^{d-1} r_t^{1-d+\theta} \int_{\epsilon/r_t}^1 du \frac{u^{-(d-\theta)}}{\sqrt{1-u^{2(d-\theta)}}} =L^{d-1} \left(\frac{\sqrt{\pi} \,\Gamma\left(\frac{1+d - \theta}{2(d-\theta)} \right)}{\Gamma\left(\frac{1}{2(d-\theta)} \right)} \,\frac{r_t^{1-d+\theta}}{1-d+\theta} -\frac{\epsilon^{1-d+\theta}}{1-d+\theta}\right)
\ee
and, from (\ref{eq:l}), the turning point is related to the length of the strip by
\be
l = r_t \int_0^1 du  \frac{u^{(d-\theta)}}{\sqrt{1-u^{2(d-\theta)}}}=\frac{\sqrt{\pi} \,\Gamma\left(\frac{1+d - \theta}{2(d-\theta)} \right)}{\Gamma\left(\frac{1}{2(d-\theta)} \right)}\,r_t\,.
\ee

Restoring $R$ and the cross-over scale $r_F$, we thus find that the entanglement entropy across a strip is
\be\label{eq:entgltheta}
\mc S = \frac{(M_{Pl}R)^d}{4(d-\theta-1)} \left[\left(\frac{\epsilon}{r_F} \right)^\theta \left(\frac{L}{\epsilon} \right)^{d-1}-\left(\frac{\sqrt{\pi} \,\Gamma\left(\frac{1+d - \theta}{2(d-\theta)} \right)}{\Gamma\left(\frac{1}{2(d-\theta)} \right)} \right)^{d-\theta} \left(\frac{l}{r_F}\right)^\theta \left(\frac{L}{l} \right)^{d-1} \right]\,,
\ee
again neglecting higher powers of $\epsilon$. Comparing with the scale invariant answer (\ref{eq:entaglAdS}), we see that the effect of the hyperscaling violation exponent is to modify the entropy by an additional power of $(\textrm{length})^\theta$. This can be understood directly in terms of scaling weights: since the metric has dimension $\theta/d$, the entanglement entropy across a $d$-dimensional region acquires a scaling weight $\theta$.

It is also useful to obtain the entanglement entropy using the method of trial surfaces described above. Choosing a cylinder in the bulk, (\ref{eq:trialA}) and (\ref{eq:stationary}) evaluate to
\be\label{eq:trialhyper}
\mc S_\text{trial} = \frac{(M_{Pl}R)^d}{4(d-\theta-1)} \left[\left(\frac{\epsilon}{r_F} \right)^\theta \frac{|\partial \Sigma|}{\epsilon^{d-1}} - \frac{1}{(d-\theta)^{d-\theta}} \left(\frac{l}{r_F}\right)^\theta \frac{|\partial \Sigma|}{l^{d-1}}\right]
\ee
where the trial value of the turning point is $r_t=(d-\theta) l$ and recall that $l=|\Sigma|/|\partial \Sigma|$ here. Eq.~(\ref{eq:trialhyper}) correctly reproduces all the physical features of (\ref{eq:entgltheta}); moreover when $\Sigma$ is a strip both expressions exactly agree for $\theta=d-1$. Eq.~(\ref{eq:trialhyper}) of course applies to general surfaces, indicating that our conclusions are valid beyond the case of strip-like region.

\subsection{Novel phases with $d-1<\theta<d$}\label{subsec:novel}

The entanglement entropy result (\ref{eq:entgltheta}) reveals interesting properties of the dual field theory. First, when $\theta=d-1$ the integral in (\ref{eq:A}) gives a logarithmic (instead of power-law) dependence, so that the entropy becomes
\be\label{eq:Sfermi}
\mc S =\frac{(M_{Pl}R)^d}{4} \left(\frac{L}{r_F}\right)^{d-1} \,\log \frac{2l}{\epsilon}\,.
\ee
The cutoff can be chosen to be of order the cross-over scale, $\epsilon \sim r_F$, where we expect the metric (\ref{eq:dsIR1}) to give a good description of the dynamics. Eq.~(\ref{eq:Sfermi}) shows a logarithmic violation of the area law, signaling the appearance of a Fermi surface in the dual theory. This case was studied in detail by~\cite{Ogawa:2011bz,Huijse:2011ef}, who identified various properties of this (strongly coupled) Fermi surface. In particular, $r_F \sim k_F^{-1}$, the inverse scale of the Fermi surface.

Another special value corresponds to $\theta=d$, where the metric becomes
\be
\label{otherspecial}
ds^2 = \frac{R^2}{r_F^2} \left(- r^{-2(z-1)} dt^2 + dr^2 + dx_i^2\right)\,,
\ee
and the geometry develops an $\mathbb R^d$ factor. In this limit, the surface that bounds the entanglement region does not move into the bulk, and the entropy is simply proportional to the volume of the entanglement region,
\be
\mc S = \frac{(M_{Pl}R)^d}{2} \,\frac{L^{d-1}l}{r_F^d}\,.
\ee
This is an extensive contribution to the entanglement entropy and suggests that the dual theory has an extensive ground state entropy. Note that the metric (\ref{otherspecial}) is not that of
$AdS_2 \times \mathbb R^d$, which shares this feature.\footnote{Similar volume laws were discussed in \cite{Barbon} (and \cite{Li:2010dr} for flat space holography) and related to nonlocality; we see no reason the metrics we are studying here are dual to non-local theories, however.}

Having understood these two limits, it is clear that in the range of parameters
\be
d-1 < \theta < d\,,
\ee
(\ref{eq:entgltheta}) predicts new violations of the area law that interpolate between the logarithmic and linear behaviors. These novel phases present various intriguing properties. To begin with, the entanglement entropy is finite: (\ref{eq:entgltheta}) does not diverge if we take the cutoff $\epsilon \to 0$. Also, in \S \ref{subsec:NEC} we learned that in general these systems have a nontrivial dynamical exponent $z > 1-\theta/d$. The correlation functions computed in \S\S \ref{sec:holography} and \ref{sec:scalars} may also provide information to further characterize these phases. For instance, for massless scalars the two-point function is $G(|\Delta x| \sim 1/|\Delta x|^{d+3 - \alpha}$, where $\alpha = \theta - (d-1)$.

In the following sections we will derive further properties of these systems by placing them at finite temperature and will comment on the possible ground states that can lead to these new phases.

\subsection{Extremizing the entanglement entropy}\label{subsec:extreme}

Finally, based on the result (\ref{eq:entgltheta}), let us determine the value of $\theta$ that extremizes the entropy. We focus on the finite contribution to $\mc S$ in the limit where the diameter is much larger than $r_F$, which is necessary in order to obtain a universal answer independent of the entanglement surface.\footnote{The same result is obtained if the metric (\ref{eq:dsIR1}) is valid all the way to $r \to 0$. In this case it is consistent to take $\epsilon \sim r_F \to 0$, and extremizing the first term of (\ref{eq:entgltheta}) with respect to $\theta$ also yields a universal answer.}

Extremizing the second term of (\ref{eq:entgltheta}) with respect to $\theta$ shows that there is a local minimum at
\be
\theta_c \approx d-1 - \mc O\left(\frac{1}{\log(l/r_F)} \right)\,.
\ee
Therefore, $\theta = d-1$ minimizes $\mc S$ in the limit of large diameter. This establishes that Fermi surfaces are local minima of the entanglement entropy.\footnote{It would be interesting to study explicitly other entanglement regions.}

We can also ask which value of $\theta$ gives the strongest $l$-dependence. Assuming $\theta \le d$, the strongest rate is a linear dependence in $l$, when $\theta=d$ (see also~\cite{Ogawa:2011bz}). The entropy then scales like the volume of the entanglement region. Intuitively, this is associated to the logarithm of the number of degrees of freedom, which scales like the volume. 

However, just from the NEC analysis in \S \ref{subsec:NEC}, we found that $\theta > d$ is allowed.\footnote{For $\theta>d$ the UV boundary is at $r=\infty$ so one has to put the entanglement region $\Sigma$ there. Eqs.~\eqref{eq:trialmass} and \eqref{eq:trialmassh} are still valid in this case whereas \eqref{eq:stationary} changes sign.} In this range, the entanglement entropy scales faster than the volume, which is not expected to correspond to a QFT behavior. It is interesting to ask whether the entanglement entropy can reveal additional properties of this regime. For this, consider again the calculation in terms of trial cylinders in the bulk. At the stationary point, the entanglement entropy (\ref{eq:trialhyper}) is a good approximation to the exact answer. The ``mass'' for fluctuations around the critical point $r_t = (d-\theta) l$ can be calculated from (\ref{eq:trialmass}), which now implies
\be\label{eq:trialmassh}
\frac{\delta^2 \mc A}{\delta r_t^2} \propto \frac{d-\theta}{r_t^2}\,.
\ee

For $\theta<d$, the stationary surface is a minimum; at $\theta=d$ the bulk surface collapses into the $r=\epsilon$ slice, explaining the extensive scaling. And for $\theta>d$ the stationary point becomes a maximum. This suggests that gravitational backgrounds with $\theta>d$ cannot appear as stable theories (at least in the regime of validity of our current analysis). \S \ref{sec:thermo} exhibits similar thermodynamic instabilities, and in the string theory realizations one finds that there is no well-defined decoupling limit. All these results suggest that theories with $\theta>d$ may not be consistent.

\section{Thermodynamics}\label{sec:thermo}

So far we studied properties of QFTs with hyperscaling violation at zero temperature, such as correlators, the energy-momentum tensor, and entanglement entropy. We will now analyze finite temperature effects in these holographic theories. After obtaining the basic thermodynamic quantities for this class of theories, we will study the entanglement entropy at finite temperature and the cross-over to the thermal result. In particular, this will reveal how the degrees of freedom responsible for entanglement are related to those that are excited by thermal effects. The reader is also referred to~\cite{Kiritsis,Sandip,Huijse:2011ef,Swingle} for related discussions. 

\subsection{Gravitational background at finite temperature}\label{subsec:finiteT}

Finite temperature effects are encoded in the general metric (\ref{eq:dsgeneral}) by introducing an emblackening factor $f(r)$ with
\be\label{eq:dsgeneralT}
ds_{d+2}^2= e^{2A(r)} \left(-e^{2B(r)} f(r) dt^2 + \frac{dr^2}{f(r)} +dx_i^2\right)\,.
\ee
The basic property of $f(r)$ is that it vanishes at some $r=r_h$; the temperature is then proportional to a power of $r_h$, as we explain in more detail below. In order to study finite temperature effects on a regime with hyperscaling violation, in the gravity side we need to take $r_F < r_h$.

Starting now from the metric (\ref{eq:dsIR1}) with hyperscaling violation, the black hole solution becomes~\cite{Sandip,Swingle}
\be\label{eq:dsIRT}
ds^2_{d+2} = \frac{R^2}{r^2} \left(\frac{r}{r_F}\right)^{2\theta/d}  \left(-r^{-2(z-1)}f(r) dt^2 + \frac{dr^2}{f(r)} + dx_i^2 \right)\,,
\ee
with
\be\label{eq:fr}
f(r) = 1 - \left(\frac{r}{r_h} \right)^{d+z-\theta}\,.
\ee
Starting from a solution with $f(r)=1$ and matter content general enough to allow for arbitrary $(z, \theta)$,\footnote{This can be accomplished for instance in Einstein gravity with a scalar and gauge field.} one can show, using the results in the Appendix, that (\ref{eq:fr}) still gives a solution. Concrete examples will be presented in \S \ref{sec:string}. As usual, the relation between the temperature and $r_h$ follows by expanding $r_h-r = u^2$, and demanding that near the horizon the metric is $ds^2 \approx du^2 + u^2 d \tau^2$, where $\tau= (2 \pi T) it$. The result is
\be\label{eq:Tz}
T= \frac{1}{4\pi} \frac{|d+z-\theta|}{r_h^z}\,.
\ee

These expressions imply that the thermal entropy, which is proportional to the area of the black hole, becomes
\be\label{eq:ST}
\mc S_T \sim (M_{Pl}R)^d  V \frac{T^{(d-\theta)/z}}{r_F^\theta}\,.
\ee
Thus, a positive specific heat imposes the condition
\be\label{eq:specificC}
 \frac{d-\theta}{z}  \ge 0 \,.
\ee
We see that the branch $\{0<z<1, \theta \ge d+z\}$ that was consistent with the NEC is thermodynamically unstable. On the other hand, $\{z \le 0, \theta \ge d\}$ is still allowed by (\ref{eq:specificC}). It would be interesting to study this case in more detail to decide whether it is consistent -- the entanglement entropy analysis of \S \ref{sec:entanglement} suggested an instability for all $\theta>d$.

Eq.~(\ref{eq:ST}) suggests that $d-\theta$ plays the role of an effective space dimensionality for the dual theory. From this point of view, $\theta=d-1$ yields a system living in one effective space dimension, i.e. a $(1+1)$-dimensional theory. Recall also that for this value of $\theta$ there is a logarithmic violation of the area law for the entanglement entropy. These points support the interpretation of $\theta=d-1$ as systems with a Fermi surface~\cite{Ogawa:2011bz,Huijse:2011ef}. The case $\theta=d$ would then correspond to a system in $(0+1)$-dimensions. In \S \ref{sec:entanglement} we found novel phases with $d-1< \theta < d$ that violate the area law. According to this interpretation, these would be systems of defects living in a fractional space dimension (between $0$ and $1$).

Notice that the behavior of the thermal entropy for $\theta=d$ can also be obtained in systems with $\theta<d$ by taking $z \to \infty$. This is the familiar $AdS_2 \times \mathbb R^{d}$ limit of a Lifshitz metric.  We see from the metrics that these systems are not equivalent, and
they are distinguished in terms of field theory observables by their correlation functions. In particular, the two-point function for a marginal operator,
\be
G(x, x') = \frac{1}{|x-x'|^{2(d+1)-\theta}}\,,
\ee
implies that positive $\theta$ increases the correlation between spatially separated points. 
Therefore, despite the extensive ground-state entropy,  $\theta=d$ does not in any sense correspond to spatially uncorrelated quantum mechanical degrees of freedom.

\subsection{Entanglement entropy and cross-over to thermal entropy}\label{subsec:entanglT}

We now study the entanglement entropy at finite temperature. This quantity is of physical interest since it illustrates how the degrees of freedom responsible for the entanglement entropy contribute to the thermal excitation. As~\cite{Swingle} argued recently, we expect a universal crossover function that interpolates between the entanglement and thermal entropies.

Finite temperature effects modify the entanglement entropy for a strip as follows:
\be\label{eq:area2T}
\mc S = \frac{M_{Pl}^d}{4} L^{d-1} \int_\epsilon^{r_t} dr\,f(r)^{-1/2}\frac{e^{d A(r)}}{\sqrt{1-e^{-2d \left(A(r) - A(r_t) \right)}}}\,,
\ee
where $r_t$ is given in terms of the length and temperature by
\be\label{eq:lT}
l = \int_0^{r_t} dr \,f(r)^{-1/2} \frac{e^{-d \left(A(r) - A(r_t) \right)}}{\sqrt{1-e^{-2d \left(A(r) - A(r_t) \right)}}}\,.
\ee
For the purpose of comparing with the thermal entropy we focus on the universal finite contribution to (\ref{eq:area2T}).

Evaluating these expressions for the metrics (\ref{eq:dsIR1}) with hyperscaling violation obtains
\be\label{eq:crossover1}
\mc S_\textrm{finite}= \frac{(M_{Pl}R)^d}{4} L^{d-1} r_t^{1+\theta-d}\, I_-\left(\frac{r_t}{r_h} \right)\;\;,\;\;l = r_t\, I_+\left(\frac{r_t}{r_h} \right)
\ee
where we have defined the integrals
\be
I_{\pm}(\alpha) = \int^1\,du\,\frac{1}{\sqrt{1-(\alpha u)^{z+d-\theta}}} \frac{u^{\pm(d-\theta)}}{\sqrt{1-u^{2(d-\theta)}}}\,.
\ee
Also, recall from (\ref{eq:Tz}) that $r_h \sim T^{-1/z}$.

We first need to express the turning point $r_t$ in terms of $l$ and $T$ and then plug this into $\mc S_{finite}$. While $I_{\pm}(\alpha)$ don't have a simple analytic form for general $\theta$, we can analyze the limits of small and large temperatures explicitly. In the small temperature regime we have $T^{1/z} l \ll 1$ or, equivalently, $r_h \gg r_t$. The lowest order thermal correction to the entanglement entropy (\ref{eq:entgltheta}) is
\be
\mc S_\textrm{finite}\propto (M_{Pl} R)^d\left(\frac{L}{l} \right)^{d-1} \left(\frac{l}{r_F}\right)^\theta \left[-\frac{1}{d-\theta-1}+ c_1(d-\theta,z) (l T^{1/z})^{d-\theta+z} + \ldots \right]
\ee
with $c_1$ a positive constant that depends on $d-\theta$ and $z$. So the finite contribution to the entanglement entropy increases by thermal effects, with a nontrivial power $T^{(d-\theta+z)/z}$. Interestingly, this dependence is in general non-analytic -- for instance, the leading thermal correction is $\sim T^{(z+1)/z}$ in systems with a Fermi surface ($\theta = d-1$). It would be interesting to understand the physical implications of these corrections.

On the other hand, in the large temperature regime, $r_h \to r_t$, and the integrals $I_\pm$ are dominated by the pole near $u=1$. Then $I_+ \approx I_- \approx l/r_h$, and expressing $r_h$ in terms of $T$ finds
\be
\mc S_\textrm{finite} \propto L^{d-1} l \,T^{(d-\theta)/z}\,,
\ee
which agrees with the thermal entropy (\ref{eq:ST}). This verifies the existence of a crossover function that interpolates between the entanglement and thermal entropy, in theories with $\theta \neq 0$.

\section{String theory realizations}\label{sec:string}

In this last section we will show how some $\theta \neq 0$ metrics arise from string theory. Theories with nontrivial hyperscaling violation have been realized so far in effective actions for Einstein, Maxwell and dilaton fields. However, this description usually breaks down in the far UV or IR, so it is important to have UV completions that explain what happens at very short or very long distances. We will accomplish this by noting that, over a wide range of scales, Dp-branes with $p \neq 3$ give rise to metrics of the form (\ref{eq:dsIR1}) with $z=1$ but $\theta \neq 0$.  This discussion has overlap with similar remarks in \cite{Eric}.

\subsection{Black-brane solutions}\label{subsec:branes}

Let us first review the necessary results on black branes in ten-dimensional supergravity. For more details and references to the literature see e.g.~\cite{Holography,Peet:2000hn}. In 10d string frame, the Dp-black brane solution is
\be
ds_{str}^2 = - \frac{f_+(\rho)}{\sqrt{f_-(\rho)}} dt^2 + \sqrt{f_-(\rho)} \sum_{i=1}^p dx_i^2 + \frac{f_-(\rho)^{-\frac{1}{2}-\frac{5-p}{7-p}}}{f_+(\rho)} d\rho^2 + \rho^2 f_-(\rho)^{\frac{1}{2}-\frac{5-p}{7-p}}d\Omega_{8-p}^2\,,
\ee
where
\be
f_{\pm}(\rho) = 1 - \frac{\rho_{\pm}^{7-p}}{\rho^{7-p}}\,.
\ee
The supergravity solution includes a dilaton
\be
e^{-2\phi(\rho)} = g_s^{-2} f_-(\rho)^{-(p-3)/2}
\ee
and RR p-form with field strength
\be
\int_{S^{8-p}} \star F_{p+2} = N\,,
\ee
where $N$ is the number of D-branes. The ADM mass is $M \propto (8-p) \rho_+^{7-p}- \rho_-^{7-p}$, and $N \propto (\rho_+ \rho_-)^{(7-p)/2}$.

It is convenient to introduce a new radial coordinate
\be
u^{7-p} = \rho^{7-p} - \rho_-^{7-p}
\ee
and define
\be
\rho_+^{7-p} = u_h^{7-p} \cosh^2 \beta\;,\;\rho_-^{7-p} = u_h^{7-p} \sinh^2 \beta\,.
\ee
Here
\be\label{eq:beta}
\sinh^2 \beta = - \frac{1}{2} + \sqrt{\frac{1}{4}+ \left(c_p g_s N (l_s/u_h)^{7-p} \right)^2}\,.
\ee
Then the metric and dilaton acquire the more familiar form
\bea
ds_{str}^2 &=& H(u)^{-1/2} \left( -f(u) dt^2 + \sum_{i=1}^p dx_i^2\right) + H(u)^{1/2} \left(\frac{du^2}{f(u)}+ u^2 d\Omega_{8-p}^2 \right)\nonumber\\
e^{\phi(u)}&=& g_s H(u)^{(3-p)/4}\,,
\eea 
with
\be
H(u) = 1 + \sinh^2 \beta\,\frac{u_h^{7-p}}{u^{7-p}}\;,\;f(u) = 1-\frac{u_h^{7-p}}{u^{7-p}}\,.
\ee

In order to compute the Bekenstein-Hawking entropy, we need to change to 10d Einstein frame. This is accomplished by rescaling
\be
ds_E^2 = (g_s^{-1} e^\phi)^{-1/2} ds_{str}^2\,.
\ee
The area of the horizon at $u=u_h$ then gives an entropy
\be
\mc S_{BH} \sim \cosh \beta\,u_h^{8-p}\,,
\ee
in 10d Planck units. This defines a thermal entropy for the dual theory on the Dp-brane, where the temperature is determined by $\cosh \beta$ and $u_h$:
\be
T \sim \frac{1}{\cosh \beta\,u_h}\,.
\ee
These results simplify in the limit of small temperature, in which case the black branes are nearly extremal. When $u_h \to 0$, (\ref{eq:beta}) gives $\cosh^2 \beta \sim \sinh^2 \beta \sim 1/u_h^{7-p}$. Note also that $\sinh^2\beta\,u_h^{7-p} \sim g_s N$ in string units. Then $T \sim u_h^{(5-p)/2} $ and
\be\label{eq:SBH}
\mc S_{BH} \sim T^{\frac{9-p}{5-p}}\,.
\ee

The (extremal) supergravity description is valid when the curvature and dilaton are small. In terms of the effective 't Hooft coupling on the branes,
\be
g_\textrm{eff}^2 = \frac{g_s N}{u^{3-p}}\,,
\ee
the solution is valid when~\cite{IMSY}
\be\label{eq:sugrarange}
1 \ll g_\textrm{eff}^2 \ll N^{\frac{4}{7-p}}\,.
\ee
At large $N$ and for $p<7$, this gives a large range of $u$ where the supergravity description can be trusted.

Notice that the dilaton grows large for $p \leq 2$ in the deep IR, and goes outside the range \eqref{eq:sugrarange}.  For example at $p=2$, the theory flows into the M-theory regime.  We emphasize that the entropy scaling \eqref{eq:SBH} is valid when the horizon is located within the regime of validity of 10D supergravity/string theory \eqref{eq:sugrarange}, and at the corresponding range of temperatures, the theory exhibits hyperscaling violation.

\subsection{$(p+2)$-dimensional effective theory and hyperscaling violation}\label{subsec:KK}

We will now compactify this theory on $S^{8-p}$ and show that it leads to hyperscaling violation. Dimensionally reducing on the sphere and changing to Einstein frame in $p+2$ dimensions obtains
\be\label{eq:effDp}
ds_{p+2}^2 = u^{(16-2p)/p} H(u)^{1/p} \left(-f(u) dt^2 + \sum_{i=1}^p dx_i^2 + H(u) \frac{du^2}{f(u)} \right)\,.
\ee
This is of the general form (\ref{eq:dsgeneral}) after the redefinition 
\be\label{eq:redef}
dr \equiv H(u)^{1/2} du\,.
\ee
Also $p=d$ in the notation of the previous sections.

Taking the near horizon limit\footnote{Note that the near horizon limit $u^{7-p}\ll g_sN$ overlaps with the supergravity range \eqref{eq:sugrarange} for $p<7$.}
of (\ref{eq:effDp}), we arrive at metrics (\ref{eq:dsIR1}) and (\ref{eq:dsIRT}) with hyperscaling violation exponent
\be\label{eq:thetap}
\theta = p - \frac{9-p}{5-p}\,,
\ee
where $r \propto u^{(p-5)/2}$. The emblackening factor $f(u)$ also reproduces the black hole solution (\ref{eq:dsIRT}). As a further check, we can compute thermal effects in this effective theory and compare with the 10d answer. For instance, plugging (\ref{eq:thetap}) into the formula for the thermal entropy (\ref{eq:ST}) indeed agrees with (\ref{eq:SBH}) for $z=1$.

We conclude that black-branes with $p \neq 3$ give rise to metrics with hyperscaling violation. This description is valid in the range of radial variables (\ref{eq:sugrarange}), and provides an explicit dual field theory realization of systems with hyperscaling violation. The field theory is given by $SU(N)$ super Yang-Mills in $(p+1)$ dimensions, with sixteen supersymmetries; for a large range of energy scales and strong 't Hooft coupling, it realizes a hyperscaling violation exponent (\ref{eq:thetap}). Notice that $\theta<0$ for $p \le 4$, and $\theta>p$ for $p \ge 6$. Of course, these values satisfy the NEC constraints found in \S \ref{subsec:NEC}.
It is important to remark that the $p \geq 6$ cases do not, however, ``decouple from gravity" and give rise to well-defined non-gravitational theories the way the $p \leq 5$ cases do.

A relevant case for condensed matter applications is that of $N$ D2-branes, which lead to $\theta=-1/3$. The string theory realization also allows us to understand the deep UV and IR limits, where the hyperscaling violating regime is not valid. The UV theory is given by the maximally supersymmetric YM theory in $2+1$ dimensions, which is asymptotically free. In the IR the theory flows to a strongly coupled conformal field theory dual to $AdS_4 \times S^7$ -- the gravity background for M2 branes.

Besides providing an explicit string theory realization of systems with hyperscaling violation, our results from the previous sections show interesting properties of maximally symmetric Yang-Mills theories in the intermediate strongly coupled regime (\ref{eq:sugrarange}). For instance, the propagator for an operator dual to a massless scalar is
\be
G(x, x')= \frac{1}{|x-x'|^{p+2-\frac{9-p}{5-p}}}\,.
\ee
The entanglement entropy is also a probe of strong dynamics. The result (\ref{eq:entgltheta}) yields
\be\label{eq:entanglDp}
\mc S \sim \frac{L^{d-1}}{\epsilon^{\frac{4}{5-p}}} - C_p \frac{L^{d-1}}{l^{\frac{4}{5-p}}}
\ee
with $C_p$ a numerical constant. The entanglement entropy for D-branes was also calculated in~\cite{Barbon}. Their interpretation of (\ref{eq:entanglDp}) was in terms of an area law and a number of degrees of freedom with nontrivial dependence on the RG scale,
\be
N_\textrm{eff} \propto \epsilon^{-\frac{(p-3)^2}{5-p}}\,.
\ee

\section{Future directions}\label{sec:conclusions}

In this paper, we have systematically analyzed the most basic holographic characteristics of the family of metrics (\ref{thetaz}), building on much earlier work.  There are many directions in
which one could imagine further developments.

On the one hand, the classes of such metrics which are known to
arise in string theory are still quite limited.  We saw here that Dp-brane metrics
in the supergravity regime, for $p \neq 3,5$, provide one class of examples.  
$AdS_2 \times \mathbb R^2$ and Lifshitz spacetimes provide another.
But the
cases of most physical interest in conventional systems, such as the $\theta = d-1$
``Fermi-surface'' - like case \cite{Huijse:2011ef}, remain to be realized.

On a related note, it would be interesting to interpret the $\theta \neq 0$ metrics
more explicitly in terms of dual field theories.  In some of the cases with $\theta < 0$, 
it may be useful to think of these metrics as simply reflecting a growth of the effective
number of degrees of freedom with temperature or energy scale -- this has been
suggested for the Dp-brane metrics in e.g. \cite{Barbon}.  On the other hand, it has
also been suggested that spin-glass phases of the random field Ising model can be
governed by nontrivial hyperscaling violation exponent \cite{FisherHuse}.  Several
groups have proposed different ways to model random or glassy phases with gravity
or D-brane constructions \cite{KKY,AllanSho,Denef,Dio}; it would be very interesting
if coarse-graining appropriately in any of these approaches, yielded metrics of the form
(\ref{thetaz}) for reasons similar to those espoused in \cite{FisherHuse}.

Also, while here we followed an effective approach studying holography on slices at finite radius (associated to the cross-over scale $r_F$), one could imagine trying to take the limit $r_F \to 0$. It would then be interesting to extend the methods of holographic renormalization~\cite{holographicRG} to metrics with hyperscaling violation.

Finally, we uncovered here some novel ``bottom up'' holographic ground states
with entanglement entropy intermediate between area law and extensive scaling, for $d-1< \theta<d$.
It is well known that systems with extensive ground state entropy, like the duals of
$AdS_2 \times \mathbb R^2$ gravity theories (or, even simpler, theories of free decoupled spins), can yield extensive entanglement entropy.  It
would be interesting to find candidate field-theoretic models which could yield the
intermediate scalings we found here.  In \cite{Huijse} and references therein, 
supersymmetric lattice models which are either ``superfrustrated'' (enjoying extensive
ground-state entropy), or frustrated with large but sub-extensive ground state
degeneracy, are described.  It is quite possible that one can construct analogous
lattice models with intermediate ground state entropies, giving rise to
entanglement scaling like that of our new phases
\cite{discussion}.  In fact, soon after this work was submitted, similar intermediate scalings of the entanglement entropy were found in field-theoretic models with impurities \cite{Huijse:2012kq}.

\section*{Acknowledgments}
We are grateful to L. Huijse, J. Polchinski, S. Sachdev, S. Shenker, E. Silverstein and B. Swingle for helpful
discussions. This research was supported in part by the US DOE under contract DE-AC02-76SF00515 and the National Science Foundation under grant no. PHY-0756174.
S.H. is supported by the ARCS Foundation, Inc. Stanford Graduate Fellowship.

\appendix

\section{Metric properties}

This sections contains results on the Ricci and Einstein tensors for the general class of metrics
\be\label{eq:dsgeneralTapp}
ds_{d+2}^2= e^{2A(r)} \left(-e^{2B(r)} f(r) dt^2 + \frac{dr^2}{f(r)} +dx_i^2\right)\,.
\ee

The nonzero elements of the Ricci tensor are
\bea
R_{tt}&=& \frac{1}{2} e^{2B(r)}f(r)\Big(\left[(d+2)A'(r) + 3B'(r)\right]f'(r)+2f(r)\left[( A'(r) +B'(r))(dA'(r)+B'(r))+ \right.  \nonumber\\
&+&  \left.A''(r)+B''(r)\right]+f''(r)\Big)\nonumber\\
R_{rr}&=&-\frac{1}{2} f(r)^{-1} \Big(\left[(d+2)A'(r)+3B'(r)\right]f'(r)+2f(r)\left[B'(r)(A'(r)+B'(r))+\right.  \nonumber\\
&+& \left.(d+1)A''(r)+B''(r)\right]+f''(r) \Big) \nonumber\\
R_{ij}&=& -\delta_{ij} \Big(df(r)A'(r)^2+f(r)A'(r)B'(r)+A'(r)f'(r)+f(r)A''(r) \Big)\,,
\eea
and the scalar curvature is
\bea
R&=&-e^{-2A(r)}\Big (2(d+1)A'(r)f'(r)+3B'(r)f'(r)+(d+1)f(r)\left[dA'(r)^2+2A'(r)B'(r)\right] \nonumber\\
&+&2f(r)\left[B'(r)^2+(d+1)A''(r)+B''(r)\right]+f''(r)\Big).
\eea
As a result, the Einstein tensor $G_{MN}= R_{MN}- \frac{1}{2} g_{MN} R$ simplifies to
\bea
G_{tt}&=&-{1\over 2}de^{2B(r)}f(r)\Big(A'(r)f'(r)+f(r)\left[(d-1)A'(r)^2+2A''(r)\right]\Big)\nonumber\\
G_{rr}&=&\frac{1}{2} f^{-1}(r) dA'(r)\Big(f(r)\left[(d+1)A'(r)+2B'(r)\right]+f'(r)\Big)\nonumber \\
G_{ij}&=& \frac{1}{2}\delta_{ij}\Big(d(d-1)f(r)A'(r)^2+2df(r)A'(r)B'(r)+2f(r)B'(r)^2+2dA'(r)f'(r)+\nonumber\\
&+&3B'(r)f'(r)+2df(r)A''(r)+2f(r)B''(r)+f''(r)\Big).
\eea

We can impose the null energy condition,
\be
T_{\mu\nu}N^\mu N^\nu\geq0,
\ee
where $N^\mu N_\mu = 0,$ on the Einstein equations $G_{\mu\nu}=T_{\mu\nu}$ to derive constraints on the metric functions $A(r), B(r), f(r).$ Choosing 
\be
N^t=\frac{1}{e^{A+B}\sqrt{f}},~~N^r=\frac{\sqrt{f}}{e^A}\cos\varphi,~~N^x=\frac{1}{e^A}\sin\varphi,
\ee
where $\theta$ is an arbitrary constant, we get the constraints
\bea
& &f(A'^2+A'B'-A'')\geq 0\nonumber\\
& &(dA'+3B')f'+2f(dA'B'+B'^2+B'')+f''\geq 0
\eea
from $\varphi=0$ and $\varphi={\pi\over 2}$ respectively. 

\section{Massive propagators for general geodesics}\label{sec:geodesics}

We consider the action 
\be
S = - m \int d r \, r^{-(d-\theta)/d}\sqrt{r^{-2(z-1)} \dot \tau^2 + \dot r^2 + \dot x^2}
\ee
where we have set $\lambda =r$ and $\tau=it$ in equation (\ref{eq:Sparticle}), this time for both $|\Delta x|$ and $|\Delta \tau|$ nonzero. The integrated $x, \tau$ equations of motion define two conserved momenta:
\bea
\Pi_x&=&\frac{r^{-(d-\theta)/d}\dot x}{\sqrt{1+\dot x^2 + r^{-2(z-1)} \dot \tau^2}},\nonumber\\
\Pi_\tau &=&\frac{r^{-(d-\theta)/d}r^{-2(z-1)}\dot \tau}{\sqrt{1+\dot x^2 + r^{-2(z-1)} \dot \tau^2}}.
\eea
We can use these to rewrite $\dot x, \dot\tau$ in terms of $r$ and $\Pi_x,\Pi_\tau$:
\be\label{eq:xdot}
\frac{dx}{dr}=\Pi_x\frac{r^{-(z-1)}}{\sqrt{r^{-2(z-\theta/d)}-\Pi_x^2r^{-2(z-1)}-\Pi_\tau^2}}
\ee
and
\be\label{eq:taudot}
\frac{d\tau}{dr}=\Pi_\tau\frac{r^{(z-1)}}{\sqrt{r^{-2(z-\theta/d)}-\Pi_x^2r^{-2(z-1)}-\Pi_\tau^2}}.
\ee
Also, using the fact that at the turning point $dr/dx, dr/d\tau=0$, we can derive a relationship between $r_t,\Pi_x,\Pi_\tau$,
\be\label{eq:constraint}
r_t^{-2(z-\theta/d)}-\Pi_x^2r_t^{-2(z-1)}-\Pi_\tau^2=0.
\ee
Plugging equations (\ref{eq:xdot}) and (\ref{eq:taudot}) back into the action, we get an expression for the total geodesic distance,
\be\label{eq:generalS}
S = - m \int d r \, \frac{r^{2\theta/d-z-1}}{\sqrt{r^{-2(z-\theta/d)}-\Pi_x^2r^{-2(z-1)}-\Pi_\tau^2}}
\ee
as a function of the conserved momenta. Equations (\ref{eq:xdot}), (\ref{eq:taudot}) and (\ref{eq:generalS}) can in principle be solved numerically for particular values of the critical exponents in order to find $|\Delta x|$, $|\Delta \tau|$, and $S$ respectively in terms of $\Pi_x$, $\Pi_\tau$ and the turning point $r_t$. These solutions, along with the constraint equation (\ref{eq:constraint}) can be used to find $S(\Delta x,\Delta\tau)$.

\newpage

\bibliographystyle{JHEP}
\renewcommand{\refname}{Bibliography}
\addcontentsline{toc}{section}{Bibliography}
\providecommand{\href}[2]{#2}\begingroup\raggedright

\end{document}